# Disseminating Research News in HCI: Perceived Hazards, How-To's, and Opportunities for Innovation

C. Estelle Smith
GroupLens Research
University of Minnesota
smit3694@umn.edu

Eduardo Nevarez
Dept. Writing Studies
University of Minnesota
nevar025@umn.edu

Haiyi Zhu
HCI Institute
Carnegie Mellon University
haiyiz@cs.cmu.edu

**ABSTRACT**
Mass media afford researchers critical opportunities to disseminate research findings and trends to the general public. Yet researchers also perceive that their work can be miscommunicated in mass media, thus generating unintended understandings of HCI research by the general public. We conduct a Grounded Theory analysis of interviews with 12 HCI researchers and find that miscommunication can occur at four origins along the socio-technical infrastructure known as the Media Production Pipeline (MPP) for science news. Results yield researchers' perceived hazards of disseminating their work through mass media, as well as strategies for fostering effective communication of research. We conclude with implications for augmenting or innovating new MPP technologies.

## 1. INTRODUCTION

Scientific research is made available for public consumption via the continually evolving "socio-technical infrastructure that supports the comprehensible dissemination of scientific results and rationale to the lay public" known as the Media Production Pipeline (MPP) [56]. A socio-technical system (STS) "considers requirements spanning hardware, software, personal, and community aspects. It applies an understanding of the social structures, roles and rights (the social sciences) to inform the design of systems that involve communities of people and technology" [21]. The MPP involves a variety of community stakeholders with different roles (i.e. researchers, communications professionals, journalists, etc.) use a variety of technologies (i.e. email, phones, databases, social media, etc.) to communicate science to the public.

The MPP is of critical importance to science and society for its role in funding science and public education. Public knowledge and perception of science affects decisions about policy and research funding for science and technology [58, 39, 19, 38, 51]. Moreover, advancements in science and technology prominently shape the future of society. Some research is considered "newsworthy" because of its relevance and relatability to the general public [26], including topics in Human-Computer Interaction (HCI) ranging from everyday activities like transportation, banking, and waste management [61] to the communal and societal effects and perceptions of algorithm-driven artificial intelligence (e.g. [14, 64, 15, 50, 57]). As these emerging technologies become more normalized and ubiquitous, both the public and policy makers need to be aware of new research, and proactive about what effects it could have on the democratic establishment [33, 40, 27].

Rather than using educational and academic venues, members of the general public still rely primarily on mass media communication to access scientific knowledge [48]. Mass media communication occurs when an organization employs a technology as a medium to communicate to a large audience [31]. However, audiences are not passive recipients of science information. Rather, their information-seeking behavior is due to motivated reasoning, framing, and political ideology based on depictions of emerging technologies that they see in media venues [17, 55]. Media studies has identified that the framing and presentation of events on mass media can systematically affect how audiences come to understand those events, making the interaction between scientists and journalists an interesting area to explore [47]. However, very little work has explored how miscommunication can be created from a socio-technical perspective. This paper aims to fill this research gap.

Moreover, scientists must publicize their findings as part of their careers, but they mostly do so through lengthy, jargon-laden publications that address other academics and remain largely inaccessible to the public. At times, scientists also use communication technologies (e.g. Twitter, blogs [56]) that in theory would connect them to general audiences. However, there is little empirical research to suggest its effectiveness, since researchers and scientists often form filter bubbles among other academics in what is known as "academic Twitter" [13]. Prior research by Vines et al. cautions HCI researchers about ongoing media engagement, which can potentially harm public understanding and perception by contributing to the miscommunication of science [61]. Supporting this sense of caution, prior work also consistently shows that miscommunication in science reporting occurs in high proportions, or even the majority of science news stories [4, 30, 44]. Yet the paradox of these warnings is that there remains a necessity for scientists and the public to communicate and engage with each other, since a lack of engagement can lead to a lack of awareness of emergent scientific knowledge. There are tensions and challenges faced by both scientists and media professionals who partner up and utilize the MPP to produce science news for the public [56, 61], but we still do not know how these challenges specifically lead to miscommunication. Therefore, from a socio-technical perspective, this research asks: *(1) How does miscommunication of HCI research occur through specific MPP mechanisms? (2) What strategies can HCI researchers use to foster effective mass communication of their work using the MPP?* Our study presents a Grounded Theory analysis of interviews with 12 HCI researchers (HCIRs) to answer these two RQs.

Our first contribution to HCI from RQ1 explores the ongoing work of how interactions mediated by an STS can result in miscommunication. We find that miscommunication can arise at four MPP origins: press releases, interviews, media incentives, and Web 2.0 affordances. We propose routes for future work to study and understand the problems at scale (sec. 5.3) or to augment current MPP processes with safeguards against miscommunication (secs. 5.1, 5.2, 5.4). Second, HCI as a discipline lacks pedagogical resources for supporting researchers to communicate their work effectively to lay audiences. Addressing RQ2, we present empirically-derived strategies to promote effective and ethical communication of research in mass media. These strategies can help scientists and policymakers to consider best practices for media engagement, and to implement pedagogical training in HCI curricula.

## 2. RELATED LITERATURE

Contemporary media theorists have argued that not all scientific results are equally salient to the general public; rather, results are selected through a "gatekeeping" mechanism [53, 61]. This "gatekeeping" is associated with agenda-setting in media organizations that emphasize certain issues over others based on features related to their "newsworthiness", including relevance, timeliness, novelty, entertainment value, humor, drama, and/or sexiness [26, 61]. Furthermore, science communication research into the accuracy of legitimate mainstream news has found that from the 1970s to the 1990s a high proportion[1] of science stories contain "errors", "misrepresentations", or "miscommunications" such as scientific or technical inaccuracies, non-scientific inaccuracies, misquotations, significant omissions, exaggerations, and distortions of emphasis [4, 44, 30]. In the following sections, we provide our working definition of scientific miscommunication, and summarize relevant work on media and communication technologies.

### 2.1 Scientific Miscommunication

The term "miscommunication" is used broadly in different contexts. For example, researchers have noted similar characteristics to "disinformation" which focuses on the intentional act of spreading misinformation in what has been called "fake news" or "post-truth" [28, 49]. This deliberate and orchestrated misinformation is spread by propaganda and agendas set to manipulate audiences for political reasons [29]. Misinformation can also be wrongly interpreted simply because some audiences are uninformed, and do not know about information that they have not been exposed to [49]. While these may all be a form of miscommunication, our working definition for this research relies on the socio-technical model of the MPP.

---
[1] Estimates range from ~20-90% of science news stories containing 1 or more errors of varying degrees of severity. We know that the problem exists, but do not have contemporary statistics available.

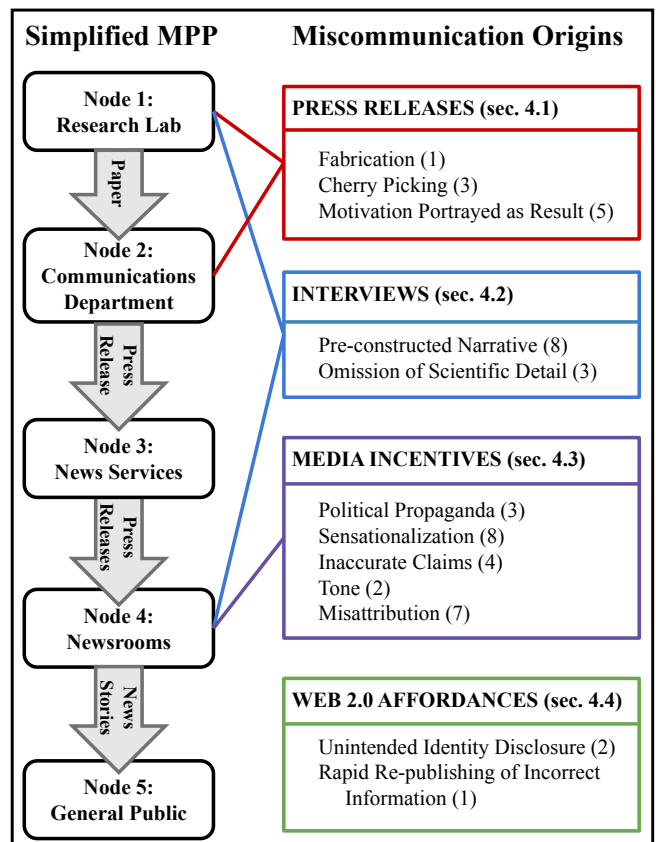

Figure 1. Overview of Miscommunication Origins along the Media Production Pipeline (MPP). Left: A simplified version of the MPP from Fig. 1 of [56]. Right: Each box corresponds to one origin of miscommunication. Lines show how each origin can be situated between nodes along the MPP. Numbers in parentheses show how many HCIRs discussed each miscommunication type.

Adapting from [4] we define: **miscommunication is the process of informing public audiences through inaccurate or misleading scientific information that is interpreted and presented by media organizations and journalists who write about science.** Our definition retains the concept of "misinterpretation" from prior work, but we are disassociating it from the concept of "intentionality." That is, deliberate "dis/misinformation" falls outside the scope of our study. Instead, our focus is on the risks of miscommunication within the MPP as a socio-technical system.

This work is essential since 73% of US adults believe the spread of inaccurate information online is a major problem with news coverage [22]. Because information can be interpreted inaccurately, it is challenging for scientists to accurately communicate emerging science to journalists [56]. One example of prior HCI work details three case studies of scientific research that was miscommunicated by media [61]. While that study focuses on public reactions to problematic news stories, the caution and critique set forth by Vines et al. are important in our work as we strive to characterize representational errors, and to describe mechanisms for how they arise.

## 2.2 Science Communication, Media, and Communication Technologies

Prior work describes the Media Production Pipeline (MPP) using five "nodes" to show how dissemination occurs: (N1) Research Labs, (N2) Communication Departments, (N3) News Services, (N4) Newsrooms, and (N5) the General Public [56]. Fig. 1 provides a very simplified overview of the MPP; see Fig. 1 of [56] for complete details. Researchers at N1 and communications professionals at N2 are described as "research advocates" who collaborate via in-person interactions, email, and phone calls to write press releases—brief summaries of newsworthy results, written in lay-accessible language and intended to get the attention of journalists. Press releases are sent to News Services at N3, which aggregate them and distribute them broadly to many Newsrooms (N4). N3 and N4 are "media outlets" which often have diverse goals that may or may not align with those of research advocates. Historically, press releases have been a major way that journalists discover stories, but new media landscapes and Web 2.0 affordances are increasingly offering new ways for journalists to find stories through social media (e.g. Facebook, Twitter), or even connect with relevant stakeholders (e.g. scientists or members of the general public) to comment on stories [8]. Thus, the MPP offers complex, varied, and non-linear routes for research advocates to engage with media outlets and the public.

While the MPP is intended to foster a helpful collaborative environment to support news dissemination, the reality is that research advocates and media outlets often have different agendas, and that MPP processes are prone to misinterpretation. For example, research advocates at N1 and N2 try to avoid language in their press releases that could be easily misinterpreted at N3 and N4 [56, 61]. As information travels through the MPP, it undergoes various interpretative processes without any risk safeguards at subsequent nodes until it finally reaches public audiences. Science communication research has identified ways in which Web 2.0 has shifted how audiences consume scientific news and how it is created; the lack of focus on the MPP in terms of miscommunication risks is our main focus.

The changing media environments from traditional print and broadcast media to exclusively online sources as the primary source for scientific information has increased the pipeline effect in which media organizations have to write more and faster stories about science [48]. It has also caused the decline of trained science journalists as cuts in news staff have led to journalists working at multiple news sections [6, 43, 18]. While big media organizations like *The New York Times, The Washington Post,* and others have been able to maintain prominent science reporters, the majority of media outlets function with minimal staff who work at multiple posts—all while increasingly incorporating new sources of data and algorithmic tools to produce news [15]. The notion of scientists filling the void of this decline is usually dismissed, as the reward systems in academic institutions do not generally encourage engagement with nonexpert audiences [48], and the lack of training in non-specialized language increases the risks of miscommunication. While this paper does not address the decline of science journalists, it does provide strategies for scientists who speak to journalists about their research in a time when there is a great need for science to be publicly disseminated.

## 3. METHODS

### 3.1 Participant Selection, Interviews, and GTM Analysis

Many studies in science communications focus on a specific scientific discipline and provide recommendations tailored to that discipline (e.g. biology [3], health [45], or climate science [4]). We chose to study Human-Computer Interaction (HCI) for several reasons. With expertise in both social and technical domains, HCI Researchers (HCIRs) are well equipped to speak to both social and technical aspects of improving MPP processes. Second, it is becoming increasingly important for HCIRs to engage in meaningful public discourse about the future of technology. While [61] offers caution about media engagement, there is no prior work that provides direct guidance for HCIRs. Finally, fewer and fewer journalists are specializing in science, and they are also under greater time and knowledge constraints. Taken together, these factors suggest that it is now important for researchers to take an active role in shaping their collaborative efforts with mass media. One limitation is that we do not have space to include perspectives from communications professionals, editors, journalists, etc. Future work should also interrogate media professionals' perspectives on socio-technical aspects of miscommunication.

This study is a secondary analysis of a subset of the same interview data presented in [56]. See [56] for more details. This paper includes interviews from 6 male and 6 female HCIRs based in the USA and UK, including 9 faculty members, 2 grad students, and 1 industry researcher. Semi-structured interviews were conducted over Skype or telephone. Participation was voluntary and uncompensated. We asked about both overall experiences and specific coverage instances with mass media, seeking to understand how their work had been communicated, how miscommunication may have arisen, and what strategies they now use to prevent miscommunication. In this subset of the [56] data, there are 8.2 hours of audio recordings, with an average interview duration of ~41 minutes.

Adopting Charmaz' approach to the Grounded Theory Method, we included inductive codes from interviews, "sensitizing" codes from prior lit, and iterated codes from later-stage analyses [11]. We transcribed and open coded interview transcripts, held team meetings to discuss/cluster codes, analyze themes, and iterate on our codebook. Finally, the first author recoded all transcripts to ensure consistent application.

### 3.2 Not-So-Anonymous Anonymity

Interviewing well-known HCIRs and subsequently publishing results to the HCI community presents a "small population"

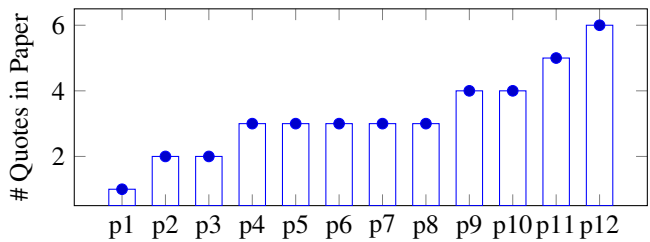

**Figure 2. Quote Distribution by Participant ID.** Of 39 total quotes, 7 (17.9%) are by 3 UK HCIRs and 32 (82.1%) are by 9 USA HCIRs.

challenge for obscuring identity; achieving perfect anonymity may not be possible [46]. All participants are already publicly known. Therefore, we do not consider them vulnerable, eliminating an ethical requirement for an extreme anonymization technique such as "un-Googling" [52]. Instead, we present rich, experiential details from participants' accounts, while maintaining confidentiality to the highest degree possible [32]. Adhering to established anonymization techniques, we omit detailed demographic information [32, 34]. Some interviewees may nonetheless be recognizable by details related to well-publicized news stories. Therefore, as an additional protocol of informed consent, we sent participants portions of the pre-submission manuscript containing their quotes. This technique enabled participants to decide for themselves about our inclusion of potentially identity-revealing details or quotes [34]. No quotes were withdrawn but several were modified to omit identifying information as a result of this protocol. We do not include participant IDs, so that even when identity can be inferred from a given quote, readers cannot tie separate quotes to the same person. In the absence of IDs, fig. 2 shows how data was selected from our sample; on average, there are 3.3 quotes/HCIR. This protocol was deemed Exempt Category 2 by the University of Minnesota IRB.

## 4. RESULTS

Participants generally felt that their own writing would be less effective at attracting lay readers; they said that journalists *"write more engagingly than scientists"* by putting an angle or spin on their work that makes the science accessible. Although most HCI Researchers (HCIRs) expressed satisfaction with some news coverage instances, all 12 participants retained concerns about at least one instance of perceived miscommunication. A senior HCIR with decades of experience said, *"One of the repeating themes I'm hearing is the need to formalize these things, not to have it just be something you need to stumble your way through, but having structure within the discipline that support the experience of working with media."* Our results provide empirical evidence that we hope can begin to provide such structure for the HCI community.

In each of the following sections, we begin by providing examples of quotes that demonstrate participants' perceptions of specific types of miscommunication of their work, along with how that miscommunication arose through a socio-technically mediated process. Fig. 1 provides an overview of four miscommunication origins along the Media Production Pipeline: (1) press releases, (2) interviews, (3) media incentives and formats, and (4) web 2.0 affordances. (See supplementary table for an expanded version of Fig. 1 that includes definitions and examples of each miscommunication type.) We do not claim that each miscommunication type can *only* arise at the below-described origins.[2] Participants (especially those with many more years of experience) often described their personal strategies for effectively mitigating miscommunication. Therefore, we conclude each section by synthesizing these strategies. We acknowledge the limitation that these strategies do not provide *specific* guidance for every situation. Thus, we recommend that researchers also work with communications professionals at their institution to develop targeted approaches for working with specific types of journalists, media, and online mediums.

### 4.1 Press Releases

Researchers usually have an opportunity to proofread press releases [56]. Nonetheless, distortions of emphasis or factual mistakes on the behalf of researchers and/or Public Information Officers are sometimes overlooked.

#### 4.1.1 Fabrication

In one case, a release posted to a university's news page had originally–but erroneously–stated that funding for a highly publicized research project came from the US Department of Defense (DOD). The error was retracted, but its original presence on the university's website generated ambiguity, and a reporter called to inquire. Although the HCIR confirmed that no money was coming from the DOD, the reporter still published that the DOD was a funder. *"The main PR guy that was helping me was like, that was shocking, because we had told him explicitly and on the record, but he went on and did it anyways."* This fabricated information consumed international news cycles for days, and may have contributed to the public perceiving the research in a more nefarious light.

However, press releases often facilitate a more subtle form of miscommunication than above. *"The problem that sometimes arises is how the press then spins a story on top of [the press release] that could be harmful, or misportrayed, or just go off in a particular way that wasn't really intended. This is where the danger zone is, even with a carefully crafted story."* For instance, participants described how journalists selected certain statements from press releases, while ignoring others.

#### 4.1.2 Cherry Picking

One HCIR described how their press release contained a list of interesting results, but the press coverage highlighted one minor result and ignored others:

> *One of the small findings was, the kids in our study said they want their parents to ask them permission before they post things about them on social media. ... We got a [feminist blog] article that's like, 'Kids Wish Parents Would Shut the Hell Up on Social Media.' That's the bit that took off. It didn't bother us because it was technically true, but it was such a minor finding that it was kind of funny that that was the sort of thing that came through.*

---

[2] In fact, some of the presented quotes were labeled with multiple codes from our codebook. We aimed to synthesize quotes to illustrate *patterns* of error causation according to participants' statements.

Another HCIR recounted how a single quote, rather than a result, was cherry-picked from the press release:

> There was one quote we left in the press release, that at the time we didn't think anything of. One thing we've noticed anecdotally is [observation about how people interact with tech]. The paper was in part motivated by that, but has a lot of other findings we wanted to highlight. ... That was the one quote that both [untrustworthy and very respectable] news organizations picked up on.

*4.1.3 Motivation Portrayed as Result*

The prior quote introduces the concept of a work's motivation, rather its results, becoming the driving force of a news piece. In HCI, press releases often mention examples of possible future technology. However, these examples, which might not yet exist, can become the primary focus of an article. One HCIR described how a motivating example from a press release was presented in international news coverage as a fully functional app. *"They went off on a tangent with one of the examples we'd given, and it just wasn't what we were doing at all. We really were just looking at the acceptability of this, and whether or not people would use it."* Nonetheless, "whether or not people would use it" became miscommunicated as a functional, complete product.

Prototypes are also a frequent research tool in HCI; depending on their fidelity, they may appear as an existing or market-ready technology. *"What really became quite apparent to us as the work was being reported more in various news articles was how, it was very quickly reconstructed, this prototype technology that we'd created, as almost a fully functional product or service, or in a sense a value proposition, something that we were pitching to customers that they might want to be using."* By focusing on the prototype, mass media failed to communicate the purpose of the research or its actual outcomes.

*4.1.4 Strategic Considerations for Press Releases*

**(1) Wait to pursue press coverage until research results and implications are mature.** When an early project has good results, it may be tempting to publicize results immediately. Yet if results are preliminary, correlational, or not yet definitive, a higher risk of miscommunication exists.

> My default is that I don't necessarily want to publicize things unless it's really very very mature, completely formed research findings, or if I know it's not dealing with something sensitive.

**(2) Evaluate every sentence in the press release cautiously with a Public Information Officer.** Even if a press release is technically accurate, miscommunication may result if a single sentence is easy to misinterpret. Motivations should be clearly delineated from results, all attributions should be accurately represented, and results should be appropriately hedged.

> A lot of it is trying to make sure that first press release is really accurate. ... It doesn't always happen, especially when on a tight time table, you know, what are the easy ways someone could misunderstand this.

One critical function of press releases is to provide journalists with a way to arrange interview(s) with author(s), however interviews can facilitate another type of miscommunication.

### 4.2 Interviews

Quotes from a press release may be undesirable for journalists to include in their stories since all newsrooms have access to those same quotes. To stand out, journalists gather unique, catchy quotes through interviews that can range from around 5 to 60+ minutes; these interviews frequently determine how the science is portrayed moreso than the original paper or press release. In most cases, journalists also condense interviews into a dramatically shorter format. *"You kinda spend a half-hour, 45 minutes, talking with someone, and then there's a one sentence summary of all that."*

*4.2.1 Pre-constructed Narrative*

Many participants discussed how journalists sometimes conduct interviews seeking to confirm a pre-constructed narrative rather than to accurately and judiciously portray research. *"They've got essentially a hypothesis, e.g., whether the use of cell phones is interfering with real life interactions...They're not trying to find out what the evidence shows, but particular scientific anecdotes consistent with that hypothesis."*

HCIRs described how reporters with a pre-constructed narrative can be very pushy. Unprepared for this, some HCIRs felt obliged to answer, and described feeling unsurprised when the resultant press coverage inappropriately connected personal identity or nationality to the science, fueled pre-existing public sentiment about certain technologies, or portrayed only results consistent with a provocative idea. For example:

> It was clear that he wanted to write a story that was like, 'Here's all the horrible things that can happen if you don't read the terms of service,' and I was like...for the most part, there's not a lot of really bad stuff hidden in here. Or even if it's worst-case-scenario, those things never happen.' But he was like, 'Tell me about the worst things that can happen,' and I was like, 'Ok,' and I knew exactly what was going to happen, which it did. The article was like, '5 Horrible Things that Can Happen if You Don't Read the Terms'...with nothing about the hedging.

This quote also demonstrates that correct results can be presented in mass media without limitations or hedging. However, journalists are often not to blame for the omission of critical scientific details.

*4.2.2 Omission of Scientific Detail*

Scientists often expect that the details of their scientific methods will be disseminated to the general public, yet the general public are rarely well enough informed or motivated to weigh competing ideas or arguments [42, 41]. Therefore, framing in journalism is a common practice (especially for science journalists) since frames organize ideas and simplify complex issues by giving some aspects greater emphasis. However, it is important for scientists to recognize and discuss which aspects of the research the journalist should consider emphasizing.

Several HCIRs expressed concerns about their own performance during interviews. Some described feeling unprepared,

or they found it difficult to explain their methods during the allotted time. Scientists need to tell a simple story, but information about factors like sample size, effect size, etc. can be critical to balance interpretation of results, and scientists may fail to adequately emphasize these details during interviews. For example, a senior HCI shared that, *"Media were talking about the manipulation without any consideration of effect size...I wouldn't say that that was deceptive because I think that we as scientists tend to not communicate notions of effect size to journalists or to the public."* Discussing and negotiating critical information and arguments in the media article needs to be collaboratively constructed by both the journalist and the scientist.

*4.2.3 Strategic Considerations for Interviews*

**(1) Engage in incentive alignment at the beginning of the interview.** Set expectations by explicitly stating one's coverage goals; ask the journalist to do the same. Both parties are thereby informed and can collaborate towards satisfactory outcomes, or recognize irreconcilable differences and disengage.

> *If I understand what a reporter individually uses to measure success, and what the newspaper uses to measure the success of an article, I can try to make sure that we're connecting up, so that I basically am giving them what they need, and then making sure it's also what I need.*

**(2) Develop a bottom line in advance.** Enter an interview with a clear idea of what must be communicated, rather than "going with the flow." Be prepared with concise, relatable statements about *why* a general audience might care, and *what* takeaways should be gleaned from the results. Be prepared to reframe questions in service of delivering prepared statements.

> *[Journalists] often have not read any of my work, it's what they get from the interview. As a result, I can emphasize and de-emphasize pieces in response. So a lot of that process is helping the journalist to ask the right questions, as opposed to them coming in with, often what I think of as kind of naive questions...A lot of what I try to do is to reframe the questions they're asking.*

**(3) Practice verbalizing accurate simplifications and analogies of scientific methods ahead of interviews.** Assume an audience that has little or no background knowledge of the domain. Eliminate jargon, but if jargon is core to the story, pre-consider how to define it simply. Clearly state how limitations affect the interpretation of results.

> *I think it's the responsibility of scientists to be able to figure out how to tell the story that's simple enough, and often that's kind of rehearsal.*

**(4) Foster an attitude of collaboration with journalists rather than of anxiety.** Improving relationship quality by investing professional trust and understanding each other's challenges can support both parties to understand and work towards mutually desirable outcomes. This collaborative engagement can be achieved in the course of a single interview. However, given a successful initial interaction, such collaboration can even be nurtured from time to time throughout the careers of both parties.

> *I think there's a mindset change we could work on, is to really enjoy speaking to the press. I think we can talk about why people dread it and how we could change it to be one of excitement and collaboration and thinking about press interviews as a type of collaboration.*

### 4.3 Media Incentives and Formats

Whereas research advocates are incentivized to communicate research accurately, the incentives of media outlets vary dramatically. For example, newsrooms can expand upon press releases to create features (e.g. thought pieces in a science-specific magazine intended to educate engaged audiences), or condense them into short attention-grabbing formats (e.g. clickbait intended primarily to generate revenue). Divergent incentives and formats can lead to miscommunication.

*4.3.1 Political Propaganda*

HCIRs discussed instances when research projects were targeted by politicians to generate anti-science public sentiment. For example, one HCIR's work was labeled with a facetious, mocking "award" for being a waste of money. *"That was a congressional staff of a congressman, whose job was to read the scientific literature, to find articles they could make fun of to support a political agenda of reducing funding."* Politicians can simply skim research grants, proceedings, or press releases, ignoring the value or results of the research in search of examples that are relatable and easy to misconstrue.

If the purpose of a news piece is simply to entertain the public, or to support a given political agenda, then the intended purpose of research advocates is easy to ignore. *"[A Morning Talk Show covered] a fairly interesting piece of work, which was that people were using social games on Facebook to do a little bit of lightweight relationship maintenance...But of course, it was represented as, 'Researchers Study Gamers. Who Cares?' It took a nuanced argument and reduced it to absurdity."*

*4.3.2 Sensationalization*

Prior literature frequently denotes sensationalization as a problematic misrepresentation of science. In our data, sensationalization was a commonly applied code, frequently co-occuring with other miscommunication types. Yet participants described mixed feelings:

> *As scientists, I think we tend to overvalue the kind of rationalist or cognitive value of these articles, but of course the popular press and the public are much more likely to approach things from an emotional point of view than a rational point of view. So it's the emotional stories that sell, not the rational ones.*

Precisely because more emotionally-oriented writing can be so effective at engaging readers, several HCIRs described that some degree of sensationalization is acceptable:

> *They're trying to find the controversy or be provocative. And I don't even necessarily fault them for that, because if they make it a little provocative and then they hold true to the points of it, and they don't over-sensationalize*

*things...I wouldn't do the work if I didn't think it was important, and so if that's what gets people to read it and to take up a conversation around it.*

Yet sensationalization may also have undesired long-term effects. *"They turned my research into clickbait...It's not inaccurate, it's just made into something that will sell. I think those practices work in the short term, but in the long term are eroding media credibility."* Thus, sensationalization is not necessarily problematic, although *over*-sensationalization is. An interesting question for future work is, where is the line, and when is it crossed?

*4.3.4 Inaccurate Claims*
Inaccurate claims sometimes co-occur with over-sensationalization. Although they may possibly arise from simple misunderstanding, *"[Reporters] often will talk with much more certainty. They'll take whatever findings and say, 'liars always talk about themselves less,' because they know that their audience wants to know, what are the ways that I can tell somebody else is lying?"* I.e. reporters may emphasize simple takeaways rather than nuanced, carefully hedged scientific claims, since that is what their audience can relate to, understand, and apply in their own lives.

As a result, scientific claims are sometimes either under- or over-reported, or over-simplified, until they are no longer accurate. HCIRs described how journalists may invert correlation as causation, or write ill-supported calls-to-action. For example, *"There were a couple [articles] that were like, 'Here, if you're depressed, this is how Instagram could help you'...I never said that people who post this content are depressed."* Others mentioned how journalists occasionally jump to conclusions and make claims that were never made in the paper, or select words that do not actually apply. For example, reporters used the wrong words to describe menstrual tracking technology. *"They're calling it 'contraception apps,' and 'ovulation tracking apps,' and that's not really what it is."*

*4.3.5 Tone*
Another instance when core research ideas may not be compromised is when the tone of news pieces diverges from the paper, press release, or interview, esp. regarding sensitive or stigmatized users. Mass media may communicate these concepts predominantly visually rather than verbally, which easily lends itself to losing the nuance of carefully chosen language:

> *The paper was talking about sensitive things, people sharing intense personal feelings, depression, suicide, self-harm, eating disorders, etc. One of the videos I saw, they were showing this young woman sitting by a fountain, scrolling down Instagram, showing random pictures and, it just, the tone of it, the feeling of the video, was not the tone of my paper.*

Or, tone may be modified to garner attention more effectively. *"He changed the tone of some of my comments to make it a lot more clickbaity...I was very careful to say that these are smart people, this is not a mental illness...that didn't come through."*

The prior quotes raise issues not only of tone, but also of how media formats introduce miscommunication. In mass media, word counts are often very low. Not only does screen space come at a premium, but human attention is also a limited resource for which there is great competition [60]. This fact can affect most other miscommunication types including, for example, misattribution.

*4.3.6 Misattribution*
Participants discussed pervasive issues with correct attribution for work, involving misspelling or omission of collaborators, university names, geographies, departments, etc. News stories can also misbalance credit for work if journalists preferentially interview and quote figures perceived to have authority by their audience. *"A lot of times, reporters want to quote the professor because apparently I have more credibility, but the students know way more than I do."* While these may be unintentional or careless errors, others may be directly related to intrinsic limitations in media formats. *"I had one that went on Wired, and it was like, my affiliation was wrong, and they wouldn't even change it. They were like, 'It's too many words.'"* This example demonstrates how mass media organizations sometimes prioritize concision and simplicity over accuracy, even in cases when errors would be trivial to resolve.

*4.3.7 Strategic Considerations for Media Incentives, Formats*

**(1) When inundated with media requests, consider the potential corrective impact of a skillfully executed interview, along with an outlet's possible incentives.** When a story goes big, researchers generally cannot answer all media requests. It can be tempting to ignore less reputable news organizations, but their readerships may be just as large. Covering the story without HCIR input may increase the potential for miscommunication.

> *People spin their own stories. When you haven't got that carefully crafted co-relationship with a media outlet, it's much harder to control.*

> *[A disreputable newspaper] emailed us. One of our co-authors was really skeptical, like [that newspaper] is crap. We spent a long time [debating about replying]. By the end of the day we responded, ok we'll do this, but by then they had already run the story without talking to us. Basically, they made up a story about our paper.*

**(2) Understand personal boundaries and pursue media formats that feel suited to your abilities.** Some participants felt that live video footage or interviews on radio or television provide great opportunities to skillfully deliver a prepared bottom line, if a researcher is adequately prepared and able to direct or re-direct conversation, whereas others only felt comfortable with interviews for written media. For example:

> *I don't do real time stuff. I don't do radio or television because there is no editing after the fact and I'm not generally articulate enough to say what I meant the first time, so it's helpful to have the longer interview format.*

**4.4 Web 2.0 Affordances**
New content generation and interaction mechanisms are increasing the volume of media available to reporters. In particular, social media posts are now routinely integrated into news stories [16], sometimes even replacing lay commentary that, before Web 2.0, had been traditionally gathered through interviews. For HCIRs in particular, this leads to a critical consideration about including public user-generated data in scientific papers.

*4.4.1 Unintended Identity Disclosure*
Because scientific papers are not broadly publicly disseminated, it is unlikely that including real user data in scientific papers will impact those users publicly. Yet when a paper gets picked up by the press, the risk of a context-switch from academic to popular audiences can, in fact, affect real users. Some HCIRs mentioned that revealing user identity may constitute an ethical breach.

One HCIR shared a story of how a piece of news impacted individual users. *"The paper had a codebook about here's how we categorize tweets, and had the tweets in there. [The journalist] searched for the tweets and then emailed the people who tweeted them and wrote the story about that."* Although HCIRs may justify including verbatim social media posts when such data are already public, this quote demonstrates that any inclusion of public data could, and occasionally does, have consequences that extend beyond the scientific community. This HCIR similarly expressed concerns about the context-shift from a scientific to public audience when news stories cover work about niche or sensitive-context online communities; if such stories go viral, they could significantly disrupt or adversely affect entire communities.

The same consideration applies to visual data. For example, an HCIR described how a journalist had directly used Instagram images presented in the paper as the visual component for their news story.

> *In the news article, they screen-shotted images included in the paper, which I had fabricated. But those images could have easily been [real users'] images ... and some people would end up with images in a news article that would just get wildly shared. ... Because of the type of work that I do, I always have an ethical considerations section [where I discuss] things like not including anything that could be traced to somebody, even if I'm looking at public data.*

*4.4.2 Rapid Re-publishing of Incorrect Information*
The number and variety of newsrooms and content-oriented businesses has grown dramatically. Many media outlets capitalize on hot stories as quickly as possible to drive heavy traffic to their sites by re-publishing press releases, entire articles, or individual bits of information from other media sites, sometimes with a different frame, emphasis, or agenda. Before the rise of Web 2.0, it would have required close to an entire workday to accomplish this. Today, journalists might iterate on breaking stories within minutes.

> *The sort of AP/Newswire stuff, that's where I see massive amounts of copy/pasting, so I don't love that. I see one article comes out, then 20 or 30, but they're basically rewrites of the original Newswire piece with slight variations...Sometimes, the initial piece would frame it as, this is a possibility, but then the follow up coverage would be...a whole day of news cycle which wasn't true.*

This quote shows that simple omissions or word changes between iterations of a story might seem insignificant, but can become like a "game of telephone." The accuracy of a story is liable to diverge further and further from the truth, the more a story is repeatedly re-published by new media outlets.

*4.4.3 Strategic Considerations for Web 2.0*

**(1) Consider the ethics of how press coverage might affect research participants or users, were results to be picked up by the press.** These considerations should occur at an early stage in the research, so that presentation of results in academic works does not unintentionally violate communities or individuals. Faculty who instruct HCI methods or ethics courses might consider lesson plans about how journalists interact with identifiable public user data presented in papers.

> *99% of the time, the only people that are reading the paper are other academics, but if suddenly, it's the world...I think that's something we should be mindful of, particularly if we're talking about small communities or vulnerable populations.*

**(2) Leverage social media independently to share results accessibly.** Some HCIRs choose to use Twitter, their own research blogs, or blogging sites (e.g. Medium.com) to disseminate their work in more broadly accessible language. Others create their own multimedia, such as YouTube videos, Massive Online Open Courses, podcasts, or infographics, to engage with the lay public or relevant stakeholders. One problem with this is the formation of academic filter bubbles, in which scientists' audience on social media is still primarily other scientists [13]. However interactions like Twitter "mentions" can amplify not only scholarly scientific impact, but also the effects of interacting with journalists, who have a much greater public following [35].

> *I wrote a blog from my research to try to translate the research myself for the public audience, because that way I knew I would have both control of it, and I would have the ability to make sure it was framed in the right way.*

**4.5 Strategies Prior to Conducting Research**
Throughout results, we have shared participants' strategies for effective engagement with media outlets when research results are available. However, participants also described important strategies worth considering far in advance of writing a paper; these considerations can enable a greater degree of control over how results are eventually disseminated.

**(1) Consider whether a proposed research project might be newsworthy to the lay public prior to completing it.** When scientific research happens to have newsworthy features such as novelty, timeliness, relevance, humor, controversy, etc., it can get picked up by the press, whether or not researchers explicitly promote it. Thus, considering its newsworthiness ahead of time promotes preparedness and strategic planning. A quick conversation with colleagues, a Public Information Officer, and/or acquaintances with little knowledge of HCI, can help researchers to gauge their own intuitions on whether or not the research might be of public interest [56].

**(2) If the research might be newsworthy, consider all dissemination route(s), especially when writing grants.** Mass media organizations tailor content to their unique audiences, which may or may not be group(s) to whom researchers believe their work is relevant. It may be more appropriate and beneficial for researchers to publicize their work independently. For instance, agencies like the National Science Foundation (NSF) in the United States require scientists to actively plan and execute the dissemination of their results. While applying for grants or awards, researchers must describe potential impact; one possible pathway to impact is by sharing results directly with relevant stakeholders. Thus, researchers should consider funding provisions to hire videographers to create public-facing videos, or to plan seminars/events attended by researchers *and* relevant lay citizens, practitioners, policy makers and/or local organizations. Engaging with mass media may be an excellent dissemination strategy when results are broadly applicable to large sectors of the public, yet interacting more directly with precisely targeted audiences, rather than disseminating massively to large audiences, allows researchers to mitigate misunderstandings in real time and gain instant feedback and critiques of their work directly from stakeholders.

> *I'm in the process of writing an application and doing my pathways to impact, and for that I have planned a dissemination event which is bringing stakeholders, wider public, policy makers, press together in an event to showcase the technology and the work that we've done, and have costed in a videographer to do some documentation.*

> *I'm doing quite a few impact seminars or policy seminars...very carefully targeted to people that are quite high up in charities or certain commercial organizations or people from central government.*

## 5. DISCUSSION

The discipline of science communications typically focuses on how audiences interact with science information. In contrast, this work examines specific socio-technical interactions that influence how information about science is created, and how those interactions can lead to miscommunication at a variety of specific origins in the Media Production Pipeline (MPP) [56] through: (1) press releases, (2) interviews, (3) media incentives/formats, and (4) web 2.0 affordances.

The shift to online media landscapes has also caused media scholarship to reassess media effect paradigms towards a preference-based model of narrowcasting information towards ideologically fragmented publics. Because of this shift, scholars are increasingly in favor of the "public engagement" model of science communications, rather than the historical "deficit model." In the deficit model, the public is viewed in terms of its *literacy*–i.e. most of the public lacks adequate scientific literacy to be able to meaningfully contribute to public discourse about science. In this view, the main goal of science communicators would be to increase the availability of information about science, and thus increase literacy. However, Bubela et al. describe the public engagement model as a practice of communicating science through deliberative exercises between scientists, journalists, and the general public that should be "an honest effort at relationship- and trust-building rather than persuasion, with mechanisms for actively incorporating the input of lay participants into decision-making" [9]. The shift towards the public engagement model is founded in the need "to facilitate the exchange of information, knowledge, perspectives, and preferences among groups that differ in expertise, power and values" [10]. This work aims to shorten gaps that exist between scientists and media organizations, and thus contributes to Wehrmann and De Bakker's call to educate and train scientists to understand the tenets of communicating their work to lay audiences, and how that contributes to the overall engagement model [63].

A shift towards the public engagement model was mirrored in our interviews, in which HCIRs discussed not only problematic representations of their work, but collaborative strategies for improving interactions with journalists or the public, especially in sections 4.2.3 (interview strategies), 4.4.3 (strategies for Web 2.0), and 4.5 (strategies prior to conducting research). For example, HCIRs suggest that embracing an attitude of collaboration with journalists, directly engaging relevant stakeholders, and budgetary provisions for public dissemination or engagement events are important ways to make their work accessible. We posit that future technological innovation could also go far to enhance mass communication of science and help build a better relationship between HCI research fields and the general public. Thus, we also offer design implications and future research directions that follow from our analysis.

### 5.1 Press Release Production
Research advocates (scientists and Public Information Officers) collaborate between Nodes 1 and 2 in the MPP to write press releases. Our research confirms that this collaboration, followed by problematic elaboration by media outlets, can contribute to miscommunication (as in [61, 45]) when they contain factual errors or supporting data and motivations that are not primary takeaways. Our prior work suggests that crowdsourcing feedback on press releases from lay stakeholders could help research advocates gauge how press releases may be interpreted by external audiences [56]; we add that new technology could also help to ensure factual accuracy and clear communication. For example, automated information-gathering and fact-checking tools might make the process of writing releases more efficient, whereas UI/UX elements of consuming press releases could be designed to shift emphasis towards key takeaways, e.g., by more explicitly denoting "Motivations," "Future Applications," "Key Takeaways," etc.

### 5.2 Tools for Interview Preparation
Scientists at Node 1 often participate in interviews with journalists at Node 4 of the MPP. Our results demonstrate specific ways in which these press interviews can result in miscommunication, either because researchers do not feel prepared, or because they may not know how to respond under pressure to potentially leading questions from journalists. Our prior work suggests that researchers could develop communities, forums, or simulation environments to facilitate discussion and learning about mass media dissemination [56]. We enrich these ideas by providing more specific guidance on *how* effective interactions can be fostered. For instance, interactive virtual spaces could help researchers practice incentive alignment with a real or simulated journalist, reframe problematic questions to deliver a prepared bottom line, and verbally rehearse their explanations simply and without jargon.

### 5.3 Coping with Problematic Dissemination

Newsrooms (Node 4 in the MPP) are driven by diverse incentives. At times, this means that newsrooms may propagate miscommunication regardless of all due diligence by research advocates. Prior work in misinformation has often used independent fact-checking websites (e.g. snopes.com) to assign a binary true/false value to online information (e.g. [62]). However, the "accuracy" of scientific communications is not typically so black and white. For instance, HCIRs describe at least 12 different scientific miscommunication types (Fig. 1) of varying severity. Our prior work suggests that future technology should "trace [news] coverage patterns and provide corrective feedback mechanisms for emergent errors" [56] without specifying: (1) what *types* of errors exist; (2) *how* they could be detected; (3) or *what type* of feedback mechanisms might be useful.

Our results in this paper address the first issue, but do not yet provide sufficient grounding to offer solutions or datasets for addressing the second and third issues. To do so, future work will require large labeled datasets, and likely new ways to collect them. We suggest that the scientific community could develop online locations similar to fact-checking websites where researchers post links to examples of miscommunicated science, labeled at the sentence- or paragraph-level with specific miscommunication categories. Such a repository could be used across the industry-academia divide, across scientific disciplines, and internationally, while providing greater affordances for both anonymity and quantitative measurement or evaluation. Furthermore, unlike [61], this work did not investigate public reactions to what researchers perceive as miscommunication. Understanding the perspectives of those who consume science news is a critical end goal, and analyzing comments on news stories is one way to capture a glimpse of how the general public view science [20, 61]. In order to facilitate studies of both researcher perceptions of the accuracy of reporting, as well as of the complex meaning-making processes used by lay citizens to make sense of news [2], such a repository should also gather users' responses to news stories.

### 5.4 Tools for Independent Media Production

Through Web 2.0, researchers at Node 1 are empowered to communicate directly with journalists at Node 4, or the public at Node 5 of the MPP. Our results demonstrate that some researchers are becoming more proactive at publicly disseminating their work, in order to increase recognition and ensure the work is accurately portrayed. However, researchers are not always effective at communicating accessibly. Our prior work suggests that crowdsourcing technology could be designed to help researchers produce multimedia independently [56]. Here, we add that such tools for researchers should instantiate a variety of potential engagement styles. For instance, templating tools could be designed to cater multimedia to different audiences, e.g. academics, practitioners (see [12]), the public broadly, or niche audiences (e.g. patients, online communities, hobbyists, etc.). When researchers aim to disseminate to the public, technology should: (1) encourage accessible narrative structures, (2) eliminate jargon, and (3) help researchers stay mindful of reading comprehension levels.

### 5.5 Limitations

Our results may be broadly useful to scientists who want to publicize their work, yet other disciplines may also face unique challenges. Future work should continue engaging with scientists from diverse disciplines, possibly by developing new infrastructures for study (sec. 5.3) to validate our results and uncover other origins of miscommunication or discipline-specific issues. Future work should also cross-validate these results with real-life scenarios, in addition to evaluating how media professionals perceive the accuracy of scientific reporting, compared to its perceived accuracy by scientists, and how it is perceived and responded to by the public.

Finally, we do not comprehensively report every possible form of miscommunication. Most notably, this work has assumed scientists' perceptions to be accurate. However, scientists have been known to spread misinformation in cases of data fabrication, falsification, or plagiarism [59]. Bad science can get published in predatory journals [5], or even reputable journals (e.g. [54], which was later retracted [37]) before breaking out into mass media (e.g. [24]). Alternatively, *questionable* science can generate widely held beliefs that may not reflect the consensus of the broader research community (e.g. [23, 25]). A recent ACM initiative suggests that peer review should hold scientists accountable to clearly state limitations and possible negative societal impacts, and that mass media reporting has far outpaced scientists in critiquing the societal impact of tech research [7]. As researchers, we must also hold ourselves accountable to present our work ethically and accurately.

### 5.6 Conclusion

Inaccurate or miscommunicated information can have devastating consequences not only for politics [1], but also for the sciences, e.g. health and biology [45, 3], climate change [4], and Human-Computer Interaction (HCI) [61, 36]. This work has qualitatively demonstrated how miscommunciation can originate along the Media Production Pipeline [56], often well before the point at which information is shareable on social media. We provide strategies for working with media outlets to reduce miscommunication. Moreover, enhancing public engagement with science and establishing more trusting relationships between scientists and the general public is a complex and multidisciplinary challenge of urgent importance [9]; future innovation in MPP technology can play an important role in these processes.

### ACKNOWLEDGEMENTS
We are grateful to our participants, to Xinyi Wang and Raghav Karumur for helping with our analysis, and to our colleagues at GroupLens Research and the anonymous reviewers whose feedback has been vital in shaping prior iterations of work. The first author acknowledges the Graduate Assistance in Areas of National Need Fellowship for funding this work.